\documentclass[useAMS,usenatbib]{mn2e}

\usepackage{amssymb,amsmath}
\usepackage{graphicx}
\usepackage{multirow}
\usepackage{upgreek}
\usepackage{natbib}

\def\cite{\citep}
\def\b50cm{50}
\usepackage{url}
\usepackage{float}
\usepackage{bm}
\footnotesize
\newdimen\minuswidth    %define @ width of minus sign for tables
\setbox0=\hbox{$-$}
\minuswidth=\wd0
\catcode`@=\active
\def@{\kern\minuswidth}
\newdimen\digitwidth    %define ! a one digit width for tables
\setbox0=\hbox{\rm0}
\digitwidth=\wd0
\catcode`!=\active
\def!{\kern\digitwidth}
\normalsize

\usepackage[usenames,dvipsnames]{color}

\footnotesize
\newdimen\digitwidth    %define ! a one digit width for tables
\setbox0=\hbox{\rm0}
\digitwidth=\wd0
\catcode`!=\active
\def!{\kern\digitwidth}
\normalsize

\title[Five new FRBs from the HTRU high latitude survey]{Five new Fast Radio Bursts from the HTRU high latitude survey: first evidence for two-component bursts}

\author[D. J. Champion et al.]
  {D.~J.~Champion$^1$, E.~Petroff$^{2,3,4}$, M.~Kramer$^{1,5}$, M.~J.~Keith$^5$, M.~Bailes$^{2,3}$, E.~D.~Barr$^{2,3}$, 
  \newauthor
  S.~D.~Bates$^{5, 6}$, N.~D.~R.~Bhat$^{2,3,7}$, M.~Burgay$^8$, S.~Burke-Spolaor$^9$, C.~M.~L.~Flynn$^{2,3}$, 
  \newauthor
   A.~Jameson$^{2,3}$, S.~Johnston$^4$, C.~Ng$^1$, L.~Levin$^{5}$, A.~Possenti$^8$, B.~W.~Stappers$^5$,
  \newauthor
  W.~van~Straten$^{2,3}$, D. Thornton,$^{5,4}$, C.~Tiburzi$^{10,1}$, A.~G.~Lyne$^5$\\
  $^1$Max-Planck-Institut f\"{u}r Radioastronomie, 
      Auf dem H\"{u}gel 69, D-53121 Bonn, Germany \\
  $^2$Centre for Astrophysics and Supercomputing, 
      Swinburne University of Technology, Mail H30, 
      PO Box 218, VIC 3122, Australia \\
  $^3$ARC Centre of Excellence for All-Sky Astronomy (CAASTRO), 
      Mail H30, Swinburne University of Technology, 
      PO Box 218, Hawthorn, \\VIC 3122, Australia \\
  $^4$CSIRO Astronomy $\&$ Space Science, 
      Australia Telescope National Facility, 
      PO Box 76, Epping, NSW 1710, Australia \\  
  $^5$Jodrell Bank Centre for Astrophysics,
      University of Manchester, Alan Turing Building,
      Oxford Road, Manchester M13 9PL, United Kingdom \\
  $^6$National Radio Astronomy Observatory, 
      PO Box 2, Green Bank, WV 24944, USA \\
  $^7$International Centre for Radio Astronomy Research,
      Curtin University, Bentley, WA 6102, Australia \\
  $^8$INAF-Osservatorio Astronomico di Cagliari,
      via della Scienza 5, 09047 Selargius, Italy\\
  $^9$NASA Jet Propulsion Laboratory, M/S 138-307, 
      Pasadena CA 91106, USA \\
  $^{10}$Fakult\"{a}t f\"{u}r Physik, Universit\"{a}t Bielefeld, Postfach 100131, D-33501 Bielefeld, Germany%\\
}
\begin{document}

\date{}

\pagerange{\pageref{firstpage}--\pageref{lastpage}} \pubyear{2015}

\maketitle

\label{firstpage}

\begin{abstract}
The detection of five new fast radio bursts (FRBs) found in the High Time Resolution Universe high latitude survey is presented. The rate implied is 6$^{+4}_{-3}\times~10^3$ (95\%) FRBs sky$^{-1}$ day$^{-1}$ above a fluence of between 0.13 and 5.9 Jy\,ms for FRBs between 0.128 and 262 ms in duration. One of these FRBs has a clear two-component profile, each component is similar to the known population of single component FRBs and are separated by 2.4(4) ms. All the FRB components appear to be unresolved following deconvolution with a scattering tail and accounting for intra-channel smearing. The two-component FRB also has the highest dispersion measure (1629\,pc\,cm$^{-3}$) of any FRB to-date. Many of the proposed models to explain FRBs use a single high energy event involving compact objects (such as neutron star mergers) and therefore cannot easily explain a two-component FRB. Models that are based on extreme versions of flaring, pulsing or orbital events however could produce multiple component profiles. The compatibility of these models and the FRB rate implied by these detections is discussed. 
\end{abstract}


\section{Introduction}
The first detected Fast Radio Burst (FRB), now known as FRB\,010724, was found in a search for pulsars using a technique to detect bright single pulses using the Parkes radio telescope \cite{lbm+07}. The burst followed the frequency-time relation associated with dispersion of light in an ionised plasma precisely but the dispersion measure (DM, the integrated free electron density along the line of sight) was more than eight times that which could be accounted for by the Milky Way. The non-repeating nature, short duration, implied extragalactic origin (due to the large DM) and therefore luminosity made it clearly different to known short duration radio transients, such as giant pulses from pulsars and rotating radio transients (RRATs).

Reprocessing of the Parkes Multibeam Pulsar Survey \cite{mlc+01} resulted in the detection of a burst very similar to FRB\,010724 \cite{kkl+11}. Again, it precisely followed the dispersion relation and was never seen to repeat but in contrast to the high Galactic latitude of FRB\,010724 this burst was only 4$^\circ$ from the Galactic plane. The dispersion measure was only 40\% above the maximum contribution from the Milky Way expected by the NE2001 model \cite{cl02}, and as this and other similar models have large uncertainties on individual lines-of-sight a Galactic origin for this burst could not be ruled out \cite{kskl12, bm14}.

With only a single bright extragalactic event, the detection of FRB\,010724 proved controversial until the discovery of four FRBs in the High Time Resolution Universe (HTRU) survey (also using the Parkes telescope) provided a first population of these events \cite{tsb+13}. These additional FRBs allowed a rate estimate of $R_{\mathrm{F\sim3\,Jy\,ms}}=1^{+0.6}_{-0.5}\times10^4$~sky$^{-1}$~day$^{-1}$. The first FRB found with a telescope other than Parkes came from the Pulsar Arecibo L-band Feed Array (P-ALFA) survey using different hardware and software \cite{sch+14}. Since then, FRB\,010125 has been found in archival data that predates the first discovery \cite{bb14}, FRB\,140514 was detected in real time allowing for fast multi-frequency follow up \cite{pbb+15} and FRB\,131104 was discovered during observations of the Carina Dwarf Spheroidal Galaxy \cite{rsj15}. Hence, a total of nine clearly extragalactic FRB detections are in the literature, but the physical model explaining them is still unknown. A reprocessing of High Time Resolution Universe survey data taken at mid Galactic latitudes ($|b|<15^\circ$, $-120^\circ<l<30^\circ$) for FRBs resulted in no detections \cite{pvj+14}. This has been shown to be 99.5\% incompatible with the rate quoted in \citet{tsb+13} 
 and suggests either a non-uniform distribution or that the detectability of FRBs varies as a function of latitude due to latitude-dependent effects of the Galaxy \cite{mj15}.

The extremely short duration of the FRBs suggests that the source must be compact and the apparent luminosity requires a coherent, energetic process ($>10^{31}$ J) \cite{tsb+13}. There are several proposed origins of FRBs including evaporating black holes \cite{Rees77}, hyperflares from soft gamma-ray repeaters \cite{pp07}, merging white dwarfs \cite{kim13} or neutron stars \cite{hl01},  collapsing supra-massive stars \cite{fr14}, supergiant pulses from pulsars \cite{cw15}, Alfv\'{e}n waves emitted from bodies orbiting a pulsar \cite{mz14}, and even cosmic string collisions \cite{cssv12}.

The long-sought after source of the ``Perytons'' observed at the Parkes telescope was found by \citet{pkb+15} to be microwave ovens. These were characterised by multi-beam detection, patchy frequency coverage etc. The events presented in this paper showed none of these properties.

In this paper we will describe the observations and analysis in Section \S\ref{sec:oanda} before describing five new fast radio bursts in Section \S\ref{sec:res}. Interestingly, for the first time, we have detected substructure in an FRB. The impact of these discoveries on the calculated rates of FRBs, and in particular the implication of the double-component FRB for proposed models of their origin is discussed in Section \S\ref{sec:disc} before we draw our conclusions in Section \S\ref{sec:conc}.

\section[]{Observations and Analysis}\label{sec:oanda}
The HTRU survey is an all-sky survey for pulsars and fast transient sources using the Parkes 64-m radio telescope in the Southern hemisphere \cite{kjv+10} and the Effelsberg 100-m telescope in the Northern \cite{bck+13}. As the pulsar population and propagation effects due to the interstellar medium depend sensitively on Galactic latitude, the survey was split into three Galactic latitudes ranges. The data presented here come from the Southern high-latitude part of the survey ($\delta<+10^\circ$) targeting fast spinning pulsars and FRBs; the pulsar search results will be presented elsewhere. This part of the survey comprised 33,500 pointings of the 13-beam receiver, each for 270 seconds with a bandwidth of 340 MHz centred at 1.3 GHz. As this is a blind search for transient events it is useful to consider the product of the field-of-view and total observing time, in this case 1549 deg$^2$~hrs calculated to the half-power beamwidths.

The publication from \citet{tsb+13} was based on processing a subset of this same survey area. At the time of that publication 316 deg$^2$ hrs had been processed with 1400 DM trials up to a maximum of 2000 pc cm$^{-3}$. A low DM cut of 100 pc cm$^{-3}$ and a minimum signal-to-noise ratio (S/N) of 9 was applied before human inspection of all candidates. Multi-beam rejection was also used to reject interference, any candidate appearing in more than nine beams was removed. This resulted in the discovery of four FRBs.

Following that analysis a new processing pipeline known as \textsc{Heimdall}\footnote{http://sourceforge.net/projects/heimdall-astro} was developed. Using GPU technology, this pipeline is considerably quicker than the previous processing while allowing a larger range of parameters to be searched \cite{kp15}. This pipeline was used to process the data presented here including, for completeness, the 316 deg$^2$ hrs in Thornton et al. The data were searched for single pulses matching a number of criteria attributed to FRBs. The search for single pulses occurs in the three dimensions of time, dispersion measure, and pulse width typical to many single pulse searches. The data were searched over widths ranging from 0.128 to 262 ms, and over 1749 DM trials from 0 to 5000 pc cm$^{-3}$. 

Individual beams of data from the receiver were searched with \textsc{Heimdall} and then run through a coincidence algorithm to identify and cluster events occurring in more than one beam. The candidates were then concatenated into a single file for the pointing. Pulses matching the following criteria were flagged as FRB candidates:

\begin{subequations}\label{eq:thresholds}
\begin{align}
	& \mathrm{S/N} \geq 10  & \\
	& \mathrm{\Delta t} \leq 2^8 \times 64 \mathrm~\upmu \mathrm{s} = 16.3~\mathrm{ms} & \\
	& \mathrm{DM}/\mathrm{DM}_\mathrm{Galaxy} > 0.9 & \\
	& N_\mathrm{beams} \leq 4 &
\end{align}
\end{subequations}

\noindent where  $\mathrm{\Delta t}$ is the pulse width, DM$_\mathrm{Galaxy}$ is the modelled Galactic DM contribution along the line of sight from NE2001 \cite{cl02}, and N$_\mathrm{beams}$ is the number of beams of the multi-beam receiver in which the signal is detected. The thresholds for this search are identical to those from \citet{pvj+14} to maintain consistency in the FRB search across the intermediate and high latitude components of the HTRU survey.

\section[]{Results}\label{sec:res}

The entire 1549 deg$^2$ hrs of the Southern HTRU high-latitude survey was processed using the \textsc{Heimdall} software. As these included data previously analysed by \citet{tsb+13}, the re-detection of the previously known FRBs served as a validation of our pipelines. Indeed, those FRBs were detected with S/Ns similar to the original processing and no new FRBs were discovered in the area previously processed. The processing of the additional 1233 deg$^2$ hrs of observations in the high-latitude survey resulted in the detection of five FRBs. As the initial processing by \citet{tsb+13} was not done in chronological order, FRB\,090625 resulted from an observation made before their publication.

Following a detection using \textsc{Heimdall}, for each FRB the full bandwidth was divided into a smaller number of (typically eight) sub-bands. A Gaussian template was convolved with a scattering tail (a one-sided exponential) using a characteristic scattering time $\tau$. The scattering time for each sub-band was related to the centre frequency of the observation by $\tau=\tau_{\mathrm{Cen}}(\nu/\nu_{\mathrm {Cen}})^{-4}$; after fitting the scattering time at the reference frequency of 1\,GHz was calculated using the same relation.
The template was a Gaussian whose width was varied to optimise the $\chi^2$-value in a least-squares fashion. For FRB\,121002 a double Gaussian was required. In all of the FRBs (including the individual components of the double FRB), the resulting widths are consistent with smearing due to intrachannel dispersion, i.e. the pulse was unresolved. Using the arrival time of the burst at reference frequency $\nu_0$ , the arrival time at a frequency $\nu$ was scaled according to a cold plasma dispersion law $t=t_0 + k\times $DM$/\nu^2$. The parameters $\tau$, DM and $t_0$ were determined in a least-squares fit using the SIMPLEX and MIGRAD algorithms from CERN's MINUIT package\footnote{http://www.cern.ch/minuit}. Uncertainties were derived using the MINUIT algorithm to explore the error matrix, which also attempts to account for correlations between parameters. An overall baseline and amplitude of the scattered pulse of each sub-band were also treated as free parameters to be fitted. The results are summarised in Table \ref{tab:frbs}.

To confirm that the apparent double peaks seen in FRB\,121002 are significant the Akaike Information Criterium was used to compare the models using a single and double Gaussian template for the FRB when dedispersed and summed across the detected bandwidth. The double Gaussian model was more likely than the single by more than 9 orders of magnitude for FRB\,121002. The same test was applied to the other FRBs in this paper. For FRB\,090625 there was no significant difference between single and double Gaussian models and for all other FRBs the single Gaussian model was clearly preferred.

FRB\,130729 was only detected in the lower half of the observing band, and it was most strongly detected at the lowest frequencies. This could be evidence of a steep spectral index but is equally consistent with the FRB coming from the edge of the beam where the receiver's sensitivity to higher frequencies diminishes quickly. The lower bandwidth makes the DM determination less precise and the detection weaker. While it is also possible that this is terrestrial radio interference the lack of a similar detection in other beams and and DM suggests that it is of astrophysical origin. Although it appear there may be a double peak structure in this FRB it is not statistically preferred.

\begin{figure}
\includegraphics[height=10cm]{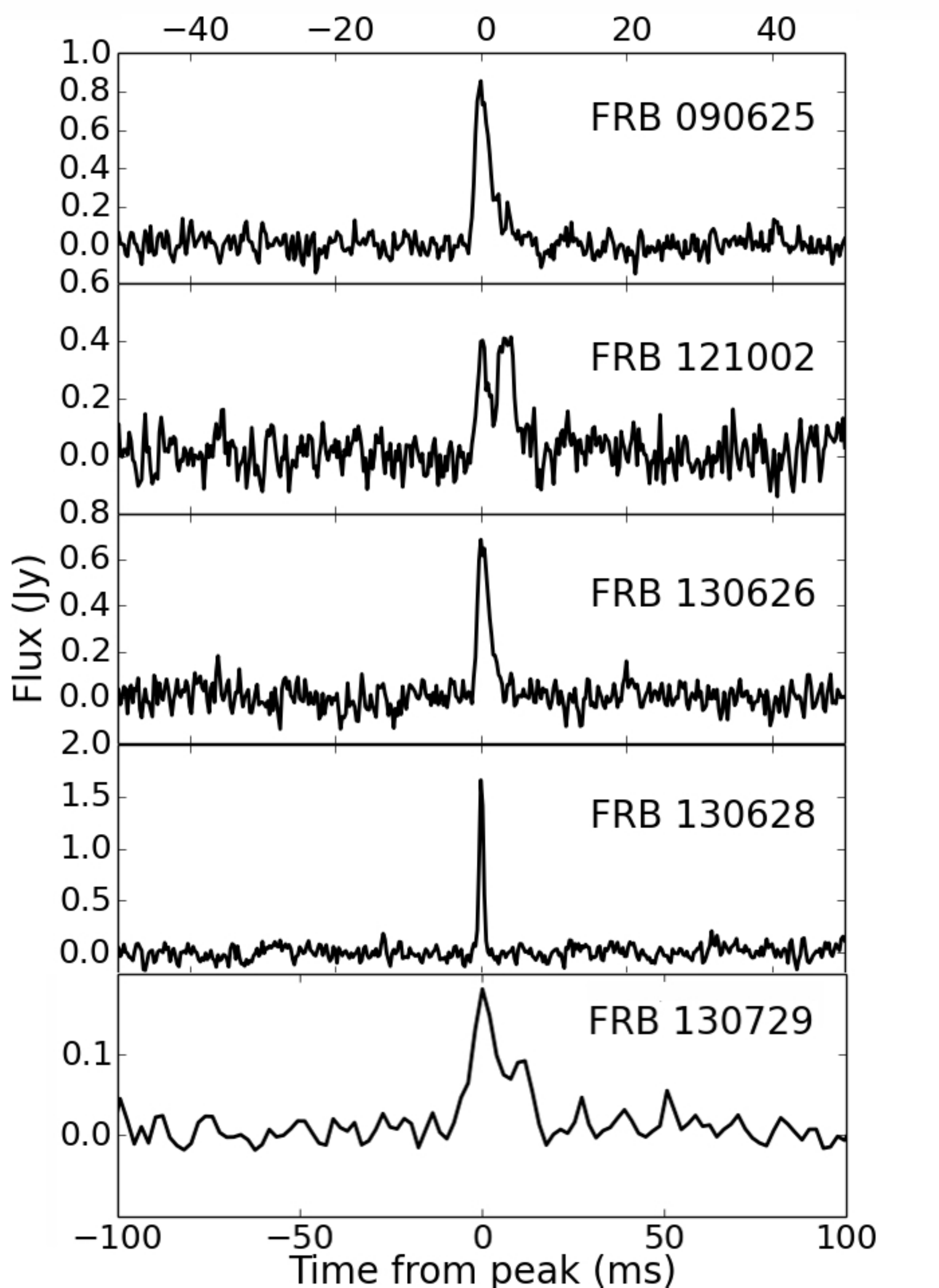}%width=9cm, 
\caption{\label{burstplot}The five FRBs presented in this work. The flux scale is calculated using the radiometer equation and assuming the FRB was at the beam centre, therefore these fluxes should be considered to be lower limits. Note that the horizontal scale for the upper four panels are at the top of the figure, while the scale for the bottom panel is below.}
\end{figure}

The new FRBs are shown in Fig. \ref{burstplot} and detailed in Table \ref{tab:frbs}. In each case the contribution to the DM from the Milky Way is estimated using the maximum value from the NE2001 model for the given line of sight. All the FRBs presented here show a very significant DM in excess of the potential Milky Way contribution. 
FRB\,121002 has the highest DM of any FRB thus far detected at 1629 pc cm$^{-3}$. If we assume that FRBs originate in an external galaxy then some of the DM will come from this host galaxy, this contribution is obviously uncertain. The intergalactic medium (IGM) DM contribution is calculated using the models of \citet{Ioka03} and \citet{Inou04}, from which we can also estimate a corresponding redshift. This gives FRB\,121002 an upper limit on redshift of $z <$ 1.3 by giving a host contribution of 0 pc cm$^{-3}$ and acknowledging that the host could contribute anything above this value depending on progenitor location and orientation (e.g. if the host galaxy is edge on). 

\begin{table*}
 \centering
% \begin{minipage}{140mm}
  \caption{The five new FRBs discovered in the HTRU high latitude survey. The time is the peak arrival time at the centre of the band. Sky positions are taken as the location of beam centre with radial errors of 7.5$'$ (the full-width half-maximum). DM$_\mathrm{Gal}$ is taken from the NE2001 model. $t_{\rm DM}$ is the intrachannel smearing time at the band centre. The scattering time $\tau$ has been scaled to a reference frequency of 1\,GHz. W$_\mathrm{Int}$ is the intrinsic width of the pulse before intrachannel smearing and scattering broaden the pulse. In the case of FRB\,121002 the scattering time and width$_\mathrm{Int}$ measurements apply to each component individually. The fluences are calculated using the radiometer equation assuming the FRB was at the beam centre and so should be considered to be lower limits.\label{tab:frbs}}
  \begin{tabular}{llllllllllll}
  \hline
 Name            & Date and Time             & \multicolumn{2}{c}{Position} & S/N & DM                  & DM$_\mathrm{Gal}$ & $t_{\rm DM}$                   & $\tau$ & W$_\mathrm{Int}$ & Fluence \\
                        &                                         & RA                 & Dec                  &                         &                                       &            &        &                                         &\\
                        &  (UTC)                            & (h:m:s)          & ($^{\circ}$:':'')  &  & \multicolumn{2}{c}{(pc\,cm$^{-3}$)}     & (ms)                 & (ms)           & (ms)                                & (Jy\,ms)\\
 \hline
 FRB\,090625 & 2009-06-25\,21:53:52.85 & \multicolumn{2}{c}{03:07:47 $-$29:55:36}  & 28 & 899.6(1)        & 32   &  1.3                 & 3.7(7)         & $<$1.9                           &$>$2.2\\
 FRB\,121002 & 2012-10-02\,13:09:18.50 & \multicolumn{2}{c}{18:14:47 $-$85:11:53}  & 16& 1629.18(2)   & 74   &  2.4                 & 6.7(7)         & $<$0.3                           &$>$2.3\\
 FRB\,130626 & 2013-06-26\,14:56:00.06 & \multicolumn{2}{c}{16:27:06 $-$07:27:48}  & 20 & 952.4(1)        & 67   &  1.4                 & 2.9(7)         & $<$0.12                         & $>$1.5\\
 FRB\,130628 & 2013-06-28\,03:58:00.02 & \multicolumn{2}{c}{09:03:02 $+$03:26:16} & 29 & 469.88(1)     & 53    &  0.7                & 1.24(7)       & $<$0.05                         & $>$1.2\\
 FRB\,130729 & 2013-07-29\,09:01:52.64 & \multicolumn{2}{c}{13:41:21 $-$05:59:43}  & 14 & 861(2)           & 31   &  1.3                 & 23(2)         &  $<$4                               & $>$3.5\\
\hline
\end{tabular}
%\end{minipage}
\end{table*}

FRB\,121002 shows a double-peaked structure. Both components can be fitted with the same DM, width and scattering time. The components have a separation of 2.4(4) ms, and the relative amplitudes of the first component to the second component (before scattering) are 0.91(2). Assuming these bursts originate beyond the Milky Way there will be a significant redshift that will have stretched the component separation between emission and detection. The redshift for FRB\,121002 is estimated to be 1.3 which results in an emission separation of 1.0(2) ms once the factor of $1+\mathrm{z}$ has been applied.
The overall width of FRB\,121002 is similar to those of FRBs 010724, 110220 and 130729 which suggests that some FRBs may have multiple components that are indistinguishable following scattering and intrachannel smearing. 

\section{Discussion}\label{sec:disc}

With the complete sample of FRBs from the high latitude survey we are able to provide an updated FRB rate with the largest sample of FRBs to date. The HTRU high latitude survey consisted of 1549 deg$^{2}$ hrs of observations in which nine FRBs were detected. Using the total time on sky and square degrees observe an all-sky rate can be calculated assuming an isotropic distribution as

\begin{equation}
9 \: \mathrm{FRBs} \times \frac{24 \: \mathrm{hrs}/\mathrm{day}\: \times\: 41253 \: \mathrm{deg}^2/\mathrm{sky}}{1549 \: \mathrm{deg}^2 \: \mathrm{hrs}} .
\end{equation}

This results in a rate of 6$^{+4}_{-3}\times~10^3$ (95\%) FRBs sky$^{-1}$ day$^{-1}$ above a fluence of between 0.13 and 5.9 Jy\,ms for FRBs between 0.128 and 262 ms in duration. While this value is lower than those previously reported in \citet{tsb+13} and \citet{sch+14}, it is consistent with these results within the 1-$\sigma$ uncertainties. It should be noted that this rate is specific to the observing setup described in this paper as this setup is not fluence-complete down to 0.9 Jy\,ms, thus it is only directly comparable to \citet{tsb+13}. The fluence-complete rate is $2.5^{+3.2}_{-1.6}\times~10^3$ (95\%) FRBs sky$^{-1}$ day$^{-1}$ above a fluence of $\sim2$ Jy\,ms \cite{kp15}.

Previous work by \citet{pvj+14} found the reported rates at high latitudes and non-detections at intermediate latitudes inconsistent with an isotropic FRB distribution with 99\% confidence. This result may be better explained with a lower overall FRB rate or a latitude-dependent rate.  \citet{mj15} suggest that diffractive scintillation at high Galactic latitudes may be aiding detection by enhancing a proportion of the underlying FRB population such that they are above our detection threshold. Using our updated rate from the full high latitude survey the probability of finding zero FRBs at intermediate latitudes assuming an isotropic distribution becomes 2.5\%. Thus the results at intermediate and high latitudes are still inconsistent with 97.5\% confidence.

For the first time an FRB has been observed that clearly shows multiple components. 
\citet{fr14} proposed that the collapse of a supra-massive star into a black hole could be the origin of FRBs. This model, building on the work in \citet{dap+13}, does predict structure within the pulse profile, specifically in the form of a leading precursor, main pulse and ringdown occurring within 1 ms. Scattering and intrachannel smearing would likely make these components indistinguishable with current observations and this timescale is significantly shorter than the component separation seen here, even when the redshift correction is applied. However the separation of the components is dependent on the rotation speed of the star, with more rapidly rotating stars having more widely separated components.

One of the models favoured by some authors relates FRBs to the giant flares from soft gamma-ray repeaters SGRs \cite[e.g.][]{pp07,tsb+13,kon+14} 
The initial gamma-ray burst usually lasts just a fraction of a second, and is then followed by hard X-ray emission with power modulated at what is thought to be the spin period of the underlying neutron star \cite[e.g.][]{hbs+05,pbg+05}. It is possible that the burst seen in gamma-rays could correspond to the initial peak. However the origin of the second peak would require structure within a single pulse or that the two bursts are two rotations of the SGR. In the latter case the spin period would be much shorter than the currently known population of SGRs. If the radio emission displayed the same dramatic decrease in flux seen in the gamma-ray emission, then no subsequent individual pulses would be expected to be seen for the known bursts. We note, that a search for periodic radio emission shortly after the burst was unsuccessful, see Chapter 6 of \citet{Thornton13} for details and a discussion of the energetics.

Another model that may be able to explain a double peaked profile was presented by \citet{cw15}. They suggest that the giant pulse behaviour of some pulsars (most notably the Crab pulsar) may extend to higher fluences. Giant pulses are generally defined as pulses that are of the order of $\sim$10 times stronger than the average fluence energy \cite{Kni07}. Giant pulses from the Crab pulsar have been seen to exceed 2\,MJy for less than 4 ns \cite{he07}. They are often extremely narrow \cite[e.g. 0.5$\upmu$s,][]{btk08} and have a power-law pulse energy distribution and in some cases are seen to show structured pulses \cite[e.g.][]{ksv10}. Giant pulses are also seen from millisecond pulsars, and the separation between the pulses could therefore represent the rotation period. Cordes et al. suggest that if this distribution continues to higher fluences then, although these ``supergiant'' pulses would be extremely rare, the volume of the Universe where they could be observed, and so the number of potential sources, makes them a possible origin of FRBs. They also suggest that for higher redshifts gravitational microlensing would play a role. In the case of the double peaked FRB two supergiant pulses at different phases of rotation of a slowly rotating pulsar, or two consecutive supergiant pulses of a fast rotating pulsar could account for the structure observed.

\citet{mz14} have suggested that a body in orbit around a pulsar could produce highly focused beams of radio emission coming from the magnetic wake of this body as it passes through the pulsar wind. They predict that such bodies would have a system of Alfv\'{e}n wings which could produce radio emission that lasts for several seconds and is composed of four pulses each with millisecond-durations. Depending upon the line-of-sight several of these pulses may be observed. While it is not yet clear if the expected rate would be, this model does predict repetition of these pulses at the orbital period of the companion. With the orbital period of the unknown companion unconstrained this cannot be ruled out. \citet{pjk+15} have ruled out periodic repeating sources with periods P $\leq$8.6~hr and sources with periods 8.6$<$P$<$21~hr at the 90 per cent confidence level.

\section{Conclusions}\label{sec:conc}
We have detected five FRBs in the southern HTRU high-latitude survey in addition to the four published by \citet{tsb+13}. The rate implied is 6$^{+4}_{-3}\times~10^3$ (95\%) FRBs sky$^{-1}$ day$^{-1}$ above a fluence of between 0.13 and 5.9 Jy\,ms for FRBs between 0.128 and 262 ms in duration, which is within the uncertainties of previously published rates. Among the new FRBs is the largest excess DM to date giving a redshift limit of $<$1.3. For the first time structure has been seen in the profile of FRBs with a two-component detection. This poses significant challenges to many of the models of FRB emission which rely on one-off high energy events. However the \citet{cw15} model of ``supergiant'' pulses and the \citet{pp07} model of hyperflares could both account for this structure. The rates expected by these models is highly uncertain and cannot at present be used as a discriminator.

\section*{Public Data Release}

The data for the five FRBs presented in this paper are made publicly available through the Swinburne gSTAR Data Sharing Cluster\footnote{TBD}. 
%All software used in this analysis is also freely available for download and use.

\section*{Acknowledgments}
The Parkes radio telescope is part of the Australia Telescope which is funded by the Commonwealth of Australia for operation as a National Facility managed by CSIRO. Parts of this research were conducted by the Australian Research Council Centre of Excellence for All-sky Astrophysics (CAASTRO), through project number CE110001020. This work used the gSTAR national facility which is funded by Swinburne and the Australian GovernmentÕs Education Investment Fund.

%\bibliographystyle{mnras}
%\bibliography{journals,myrefs,modrefs,psrrefs,crossrefs} 

\label{lastpage}

\end{document}